\documentclass[9pt,twocolumn]{extarticle}
\pdfoutput=1
\usepackage{soul}
\usepackage{titlesec} 

\usepackage[superscript, nomove]{cite}

\usepackage{float}
\usepackage{caption}
\usepackage{lipsum,ulem}
\usepackage[verbose]{placeins}

\usepackage{dsfont}
\usepackage{graphicx}
\usepackage{subcaption}
\usepackage[margin=0.9in]{geometry}
\usepackage[usenames,dvipsnames]{xcolor}
\usepackage[colorlinks,linkcolor=Blue,urlcolor=Blue,citecolor=Blue]{hyperref}
\usepackage{amsmath,amssymb}
\usepackage[squaren]{SIunits}

\usepackage{mathpazo}
\usepackage{courier}
\normalfont
\usepackage[T1]{fontenc}
\usepackage{caption}
\usepackage{upgreek}

\newcommand{\ad}[1]{\textsuperscript{#1}\kern-2pt}

\makeatletter
\def\blx@maxline{77}
\makeatother

\usepackage{capt-of}

\def\mytitle{In-situ operation of amorphous circuits under heavy-ion irradiation } 

\title{\vspace{-1.0cm}\Huge\textbf{\textrm{\mytitle}}}  

\author{Xuanzhe Sha,$^{1,2*}$ Shun Liao,$^{1,2,3*}$ Xiaoxi Li,$^{1,2,3*}$ Chengyuan Li,$^{4,5}$ Jianli Liu,$^{6}$ Yu Pan,$^{7}$ Wenhai Wang,$^{8}$\\  Yu Ye,$^{9,10}$ Chengxin Zhao,$^{11\dagger}$ Liyi Li,$^{4\dagger}$ Hanwen Wang,$^{3\dagger}$ Zheng Vitto Han,$^{1,2,3\dagger}$ Jianming Lu$^{3\dagger}$}

\date{} 
\begin{document}
\twocolumn[{
\maketitle 
\vspace{-5mm}
\begin{center}
\begin{minipage}{1\textwidth}
\begin{center}
\textit{
\\\textsuperscript{1} State Key Laboratory of Quantum Optics Technologies and Devices, Institute of Opto-Electronics, Shanxi University, Taiyuan 030006, China
\\\textsuperscript{2} Collaborative Innovation Center of Extreme Optics, Shanxi University, Taiyuan 030006, China
\\\textsuperscript{3}Liaoning Academy of Materials, Shenyang 110167, China
\\\textsuperscript{4} Shenyang National Laboratory for Materials Science, Institute of Metal Research, Chinese Academy of Sciences, Shenyang 110016, China
\\\textsuperscript{5} School of Material Science and Engineering, University of Science and Technology of China, Anhui 230026, China
\\\textsuperscript{6}State Key Laboratory of Space Environment Interation with Matters, Harbin Institute of Technology, China
\\\textsuperscript{7}School of Science, Tianjin University, Tianjin 300350, China
\\\textsuperscript{8}College of Electrical Engineering, Hebei University of Architecture, Zhangjiakou, 075000, China
\\\textsuperscript{9}Collaborative Innovation Center of Quantum Matter, Beijing, China
\\\textsuperscript{10} State Key Lab for Mesoscopic Physics and Frontiers Science Center for Nano-Optoelectronics, School of Physics, Peking University, Beijing, China
\\\textsuperscript{11}School of Integrated Circuits, Jiangnan University, Wuxi 214401, China
\vspace{5mm}
\\{$\dagger$} Corresponding to: chengxin.zhao@jiangnan.edu.cn; liliyi@hit.edu.cn; hwwang@lam.ln.cn vitto.han@gmail.com; jmlu@lam.ln.cn
\\{$\star$} These authors contribute equally.
\vspace{5mm}
}
\end{center}
\end{minipage}
\end{center}

\setlength\parindent{13pt}
\begin{quotation}
\noindent 
\section*{Abstract}
{\textbf{Radiation-hardened electronics using semiconductors beyond silicon are essential for computation and control in extreme environments. Yet complex digital circuits based on such material platforms operating $\textit{in situ}$ under heavy-ion irradiation remain largely unexplored. Here, we show a timing circuit based on amorphous thin-film semiconductors at the 100-transistor scale, and demonstrate its robust operation through a functional “Hello World” ASCII output sequence. Beyond static device characterization, we evaluate the circuit under powered heavy-ion irradiation using tantalum ions, providing an operationally relevant assessment of radiation tolerance at the system level. Under a high particle flux of $2.5\times10^3$~ions cm$^{-2}$ s$^{-1}$, the circuit maintains stable operation during the irradiation test, achieving a total fluence of $1\times10^6$~ions cm$^{-2}$, establishing a milestone of prolonged powered digital operation under extreme conditions. Our work expands the design space of radiation-tolerant electronics, highlighting amorphous semiconductors as a promising foundation for digital circuits deployed in harsh environments.}}
\end{quotation}
}]

\newpage 
\clearpage
%Radiation-hardened, novel semiconducting-materials-based electronics are essential for expanding the capability of computation and control in space, nuclear, and other extreme environments. Yet demonstrations of complex digital operation under simultaneous electrical bias and heavy-ion exposure remain scarce, particularly for large-area semiconductor platforms beyond conventional silicon technologies. 

%These results show that amorphous thin-film semiconductor electronics can support non-trivial sequential logic with substantial radiation resilience.

\section*{Introduction}

Electronics capable of operating in radiation-rich environments are essential for a broad range of technologies, including space systems, nuclear instrumentation, high-energy physics infrastructure and autonomous platforms deployed in harsh environments.\cite{prinzie2021low,vogl2019radiation} In such settings, energetic particles can induce transient faults, accumulated ionization damage and catastrophic functional failure, posing a major challenge for reliable computation, control and communication.\cite{baba2024radiation, barnaby2006total, schwank2008radiation, srour2013displacement} To address these issues, radiation resilience has traditionally been pursued through materials optimization, device engineering and system-level protection strategies.\cite{kannaujiya2025radiation, muhammad2023radiation, chatzikyriakou2018total,wang2022ultralow} In particular, redundancy-based circuit architectures and heavy shielding have been widely adopted to suppress or tolerate radiation-induced errors. However, these approaches often incur substantial penalties in footprint, weight, complexity and power consumption, which become increasingly restrictive for compact, lightweight and distributed electronic systems.\cite{lacoe2008improving, hasanbegovic2016heavy} Developing intrinsically radiation-resilient semiconductor platforms that can sustain functional circuit operation under irradiation is therefore an important goal for next-generation harsh-environment electronics.\cite{zhu2026radiation, zhu2020radiation, schranghamer2023radiation, hu2023true}

Amorphous oxide semiconductors, particularly In--Ga--Zn-based systems, provide an attractive materials platform in this context. Owing to their wide bandgap, low-temperature processability, large-area uniformity and compatibility with thin-film integration, these materials have been extensively explored for displays, sensors and emerging integrated electronics.\cite{bao2025amorphous, biggs2021natively, ozer2024bendable} Yet their potential for radiation-resilient digital circuits remains largely underexplored, especially beyond the level of individual transistors. More broadly, most prior efforts in radiation-tolerant electronics have focused on crystalline semiconductor technologies and on static device metrics\cite{pearton2023radiation,zhang2025large, zhou2025hole,arnold2019extraordinary}, whereas demonstrations of non-trivial sequential logic under simultaneous electrical bias and heavy-ion irradiation remain scarce. An especially intriguing opportunity arises from the ultrathin-body geometry accessible in oxide thin-film transistors: when the active semiconducting layer is confined to only a few nanometres, the interaction volume available for charge deposition by energetic ions can be drastically reduced. This suggests a distinct route to radiation resilience that does not rely primarily on circuit redundancy or external shielding, but instead leverages the geometric and materials characteristics of the semiconductor itself.

Here we demonstrate radiation-resilient digital timing circuits based on ultrathin amorphous InGaZnO (IGZO) thin-film transistors. We first establish the transistor and inverter characteristics required for sequential logic and implement a master--slave D-type flip-flop as the fundamental timing unit. We then integrate these building blocks into cascaded parallel-register circuits and realize an 8-bit timing system capable of generating the ASCII-coded string ``HELLO WORLD''. Under powered heavy-ion irradiation using $^{181}$Ta ions, the circuits retain stable flip-flop operation and sustain multi-bit ASCII output at a flux of $2.5\times10^{3}~$$\mathrm{ions\ cm^{-2}\ s^{-1}}$,  with an exposure of a total fluence of $10^{6}$ ions/cm$^{2}$. These results establish a system-level demonstration of functional amorphous thin-film digital electronics under \textit{in situ} heavy-ion exposure, and point to ultrathin oxide semiconductors as a promising platform for lightweight and scalable radiation-resilient integrated circuits.

\section*{Results and Discussion}
\noindent\textbf{Fabrications and characterizations of IGZO FETs.} 
\\
Amorphous IGZO thin films were deposited by magnetron sputtering and used as the active semiconductor platform for the digital timing circuits investigated in this work. The use of an amorphous oxide semiconductor is motivated by its wafer-scale uniformity, low thermal budget, and compatibility with large-area device integration, while preserving transistor characteristics suitable for sequential logic operation.\cite{geng2023thin} Thin-film transistors were subsequently fabricated through standard microfabrication processes, including channel definition, gate dielectric integration, source/drain electrode formation, and interconnect patterning to realize circuit-level building blocks. Full details of thin-film growth, device fabrication, and circuit processing are provided in the Methods section, and the sample preparation sequence is summarized in Supplementary Fig. 1. Here to mitigate linear energy transfer (LET) effects, an ultrathin channel with a thickness of approximately 2 nm was employed (Supplementary Fig. 2).

Figure~1 introduces the device concept and the circuit platform used in this study. In conventional strategies for radiation-resilient electronics, one common route is circuit-level redundancy, as schematically illustrated in Fig.~1a, where multiple parallel elements are combined through a voter architecture to suppress radiation-induced failure in an individual unit.\cite{lyons1962use,samudrala2004selective,sterpone2005analysis} A second established route is package-level shielding, shown in Fig. 1b, in which a heavy-metal encapsulation layer is employed to attenuate incoming energetic particles before they reach the active electronics. \cite{daneshvar2021multilayer,mangeret1996effects} Although both approaches are effective, they typically introduce substantial penalties in circuit footprint, system complexity, weight, and power consumption. Figure~1c outlines the alternative design philosophy explored here: the use of an ultrathin semiconductor body, exemplified by a nanometre-scale amorphous oxide channel, to intrinsically minimize the interaction volume available for heavy-ion energy deposition. In such a geometry, the active semiconducting region is confined to only a few nanometres, substantially reducing the material thickness through which ionizing particles can deposit charge.

To evaluate this concept at the circuit level, we designed and fabricated a D-type flip-flop (DFF), which serves as a representative digital timing element. An optical micrograph of the fabricated circuit is shown in Fig. 1d, and the corresponding circuit schematic is presented in Fig. 1e. The DFF adopts a master-slave architecture composed of cascaded clocked inverters and transmission-gated switching elements, enabling edge-controlled storage and transfer of logic states.\cite{yu2016design} This structure provides a compact yet functionally meaningful platform for assessing whether amorphous thin-film transistors can support stable sequential logic operation before, and ultimately during irradiation.

 \begin{figure*}[ht!]
 	\centering
 	\includegraphics[width=0.86\linewidth]{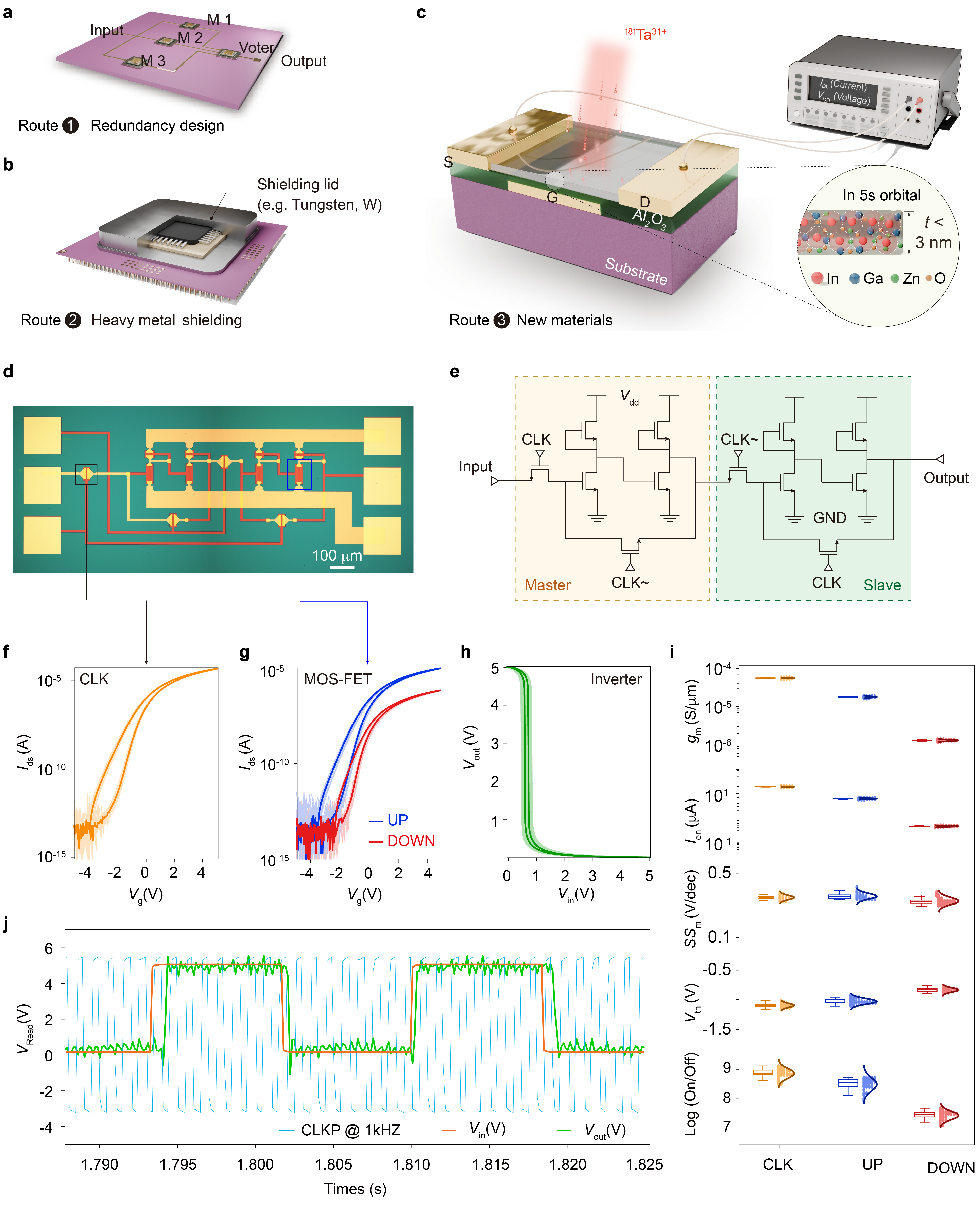}
 	\caption{
    \textbf{Design concept, device platform and circuit validation of amorphous IGZO timing circuits.} \textbf{a}, Schematic illustration of a redundancy-based design strategy for radiation-resilient circuits, in which parallel logic units are combined through a voter architecture to mitigate single-unit failure. \textbf{b}, Schematic illustration of a shielding-based strategy, where heavy-metal packaging is used to attenuate incoming radiation before it reaches the active circuit. \textbf{c}, Schematic of the ultrathin-semiconductor route explored in this work, in which the active amorphous semiconductor channel is confined to a nanometre-scale thickness, thereby minimizing the interaction volume for heavy-ion energy deposition. \textbf{d}, Optical micrograph of the fabricated D-type flip-flop (DFF) circuit based on amorphous IGZO thin-film transistors. Scale bar, 100 $\mu$m. \textbf{e}, Circuit diagram of the master-slave DFF used as the representative timing circuit in this study. \textbf{f}, Transfer characteristics of the transistor used in the clock input path. \textbf{g}, Transfer characteristics of representative IGZO MOS-FETs used in the inverter stages of the DFF, measured for different channel-width-to-length ratios. \textbf{h}, Voltage transfer curve of a typical inverter constructed from the IGZO transistors, showing clear logic inversion and sharp switching behaviour. \textbf{i}, Statistical distribution of threshold voltage extracted from multiple IGZO transistors. \textbf{j}, Dynamic operation of the DFF measured at a clock frequency of 1 kHz, demonstrating stable sequential switching between input and output logic states.
}
	\label{fig:fig1}
 \end{figure*}

 \begin{figure*}[ht!]
 	\centering
 	\includegraphics[width=0.92\linewidth]{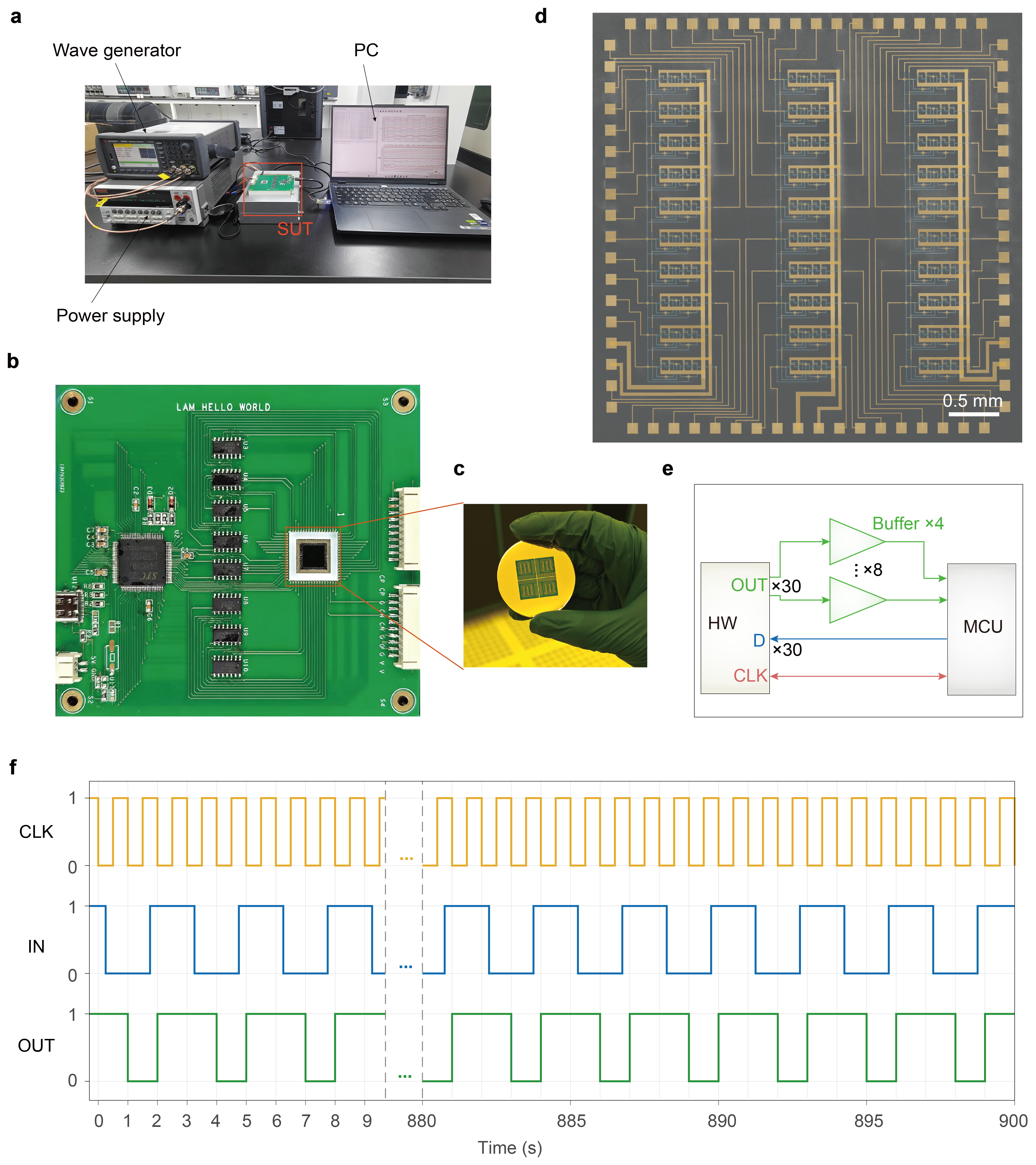}
    \caption{\textbf{Experimental platform and cascaded DFF circuits for sequential digital output.} \textbf{a}, Photograph of the electrical measurement setup used to test the amorphous thin-film timing circuits, including the wave generator, power supply, sample under test and computer-controlled readout. \textbf{b}, Photograph of the custom printed circuit board designed for interfacing and testing the timing-circuit chip. \textbf{c}, Photograph of the fabricated amorphous thin-film chip mounted for measurement. \textbf{d}, Optical micrograph of a representative cascaded DFF array forming a parallel-register circuit. Scale bar, 1000 $\mu$m. \textbf{e}, Schematic diagram of the measurement and readout architecture, showing the hardware timing circuit, buffer stages and microcontroller unit (MCU). \textbf{f}, Representative dynamic waveform of a D-type flip-flop, showing the input, clock and output signals, and confirming correct sequential latching operation.
 	}
    \label{fig:fig2}
 \end{figure*}

\begin{figure*}[ht!]
 	\centering
 	\includegraphics[width=0.94\linewidth]{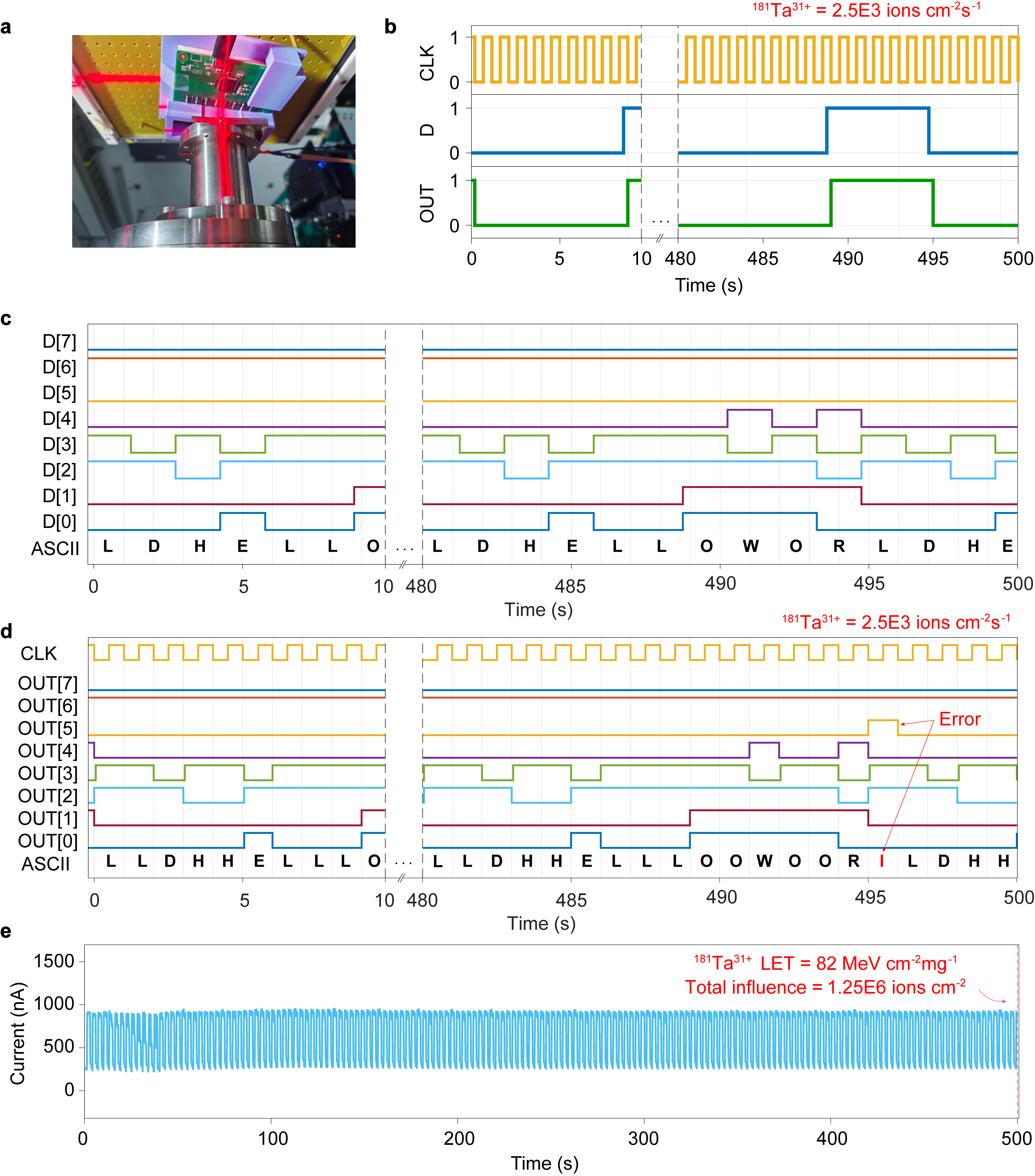} 
    \caption{\textbf{Powered heavy-ion irradiation test of amorphous thin-film timing circuits.} \textbf{a}, Photograph of the \textit{in situ} heavy-ion irradiation experiment under electrical bias. The red laser beam was used only for sample alignment and was switched off during irradiation and measurement. \textbf{b}, Representative input, clock and output waveforms of a D-type flip-flop measured during continuous irradiation by $^{181}$Ta ions at a flux of $2.5\times10^{3}~\mathrm{s^{-1}}$, showing stable operation over 500~s. \textbf{c} input and \textbf{d} output of 8-bit ASCII sequence of the cascaded timing circuit under non-irradiated operation, reproducing the programmed string ``HELLO WORLD''. The timing circuits were measured during continuous powered irradiation by $^{181}$Ta ions at a flux of $2.5\times10^{3}~\mathrm{s^{-1}}$. The output remains largely correct over 500 s, with a single erroneous code marked by the red arrow. \textbf{e}, Operating current of the timing circuit as a function of time during continuous irradiation, showing stable periodic switching up to a total ionizing dose of $10^{6}$ under $^{181}$Ta irradiation with a linear energy transfer of $82~\mathrm{MeV\,cm^{2}\,mg^{-1}}$.
}
 	\label{fig:fig3}
 \end{figure*}

  \begin{figure}[ht!]
  \centering
 	\includegraphics[width=0.9\linewidth]{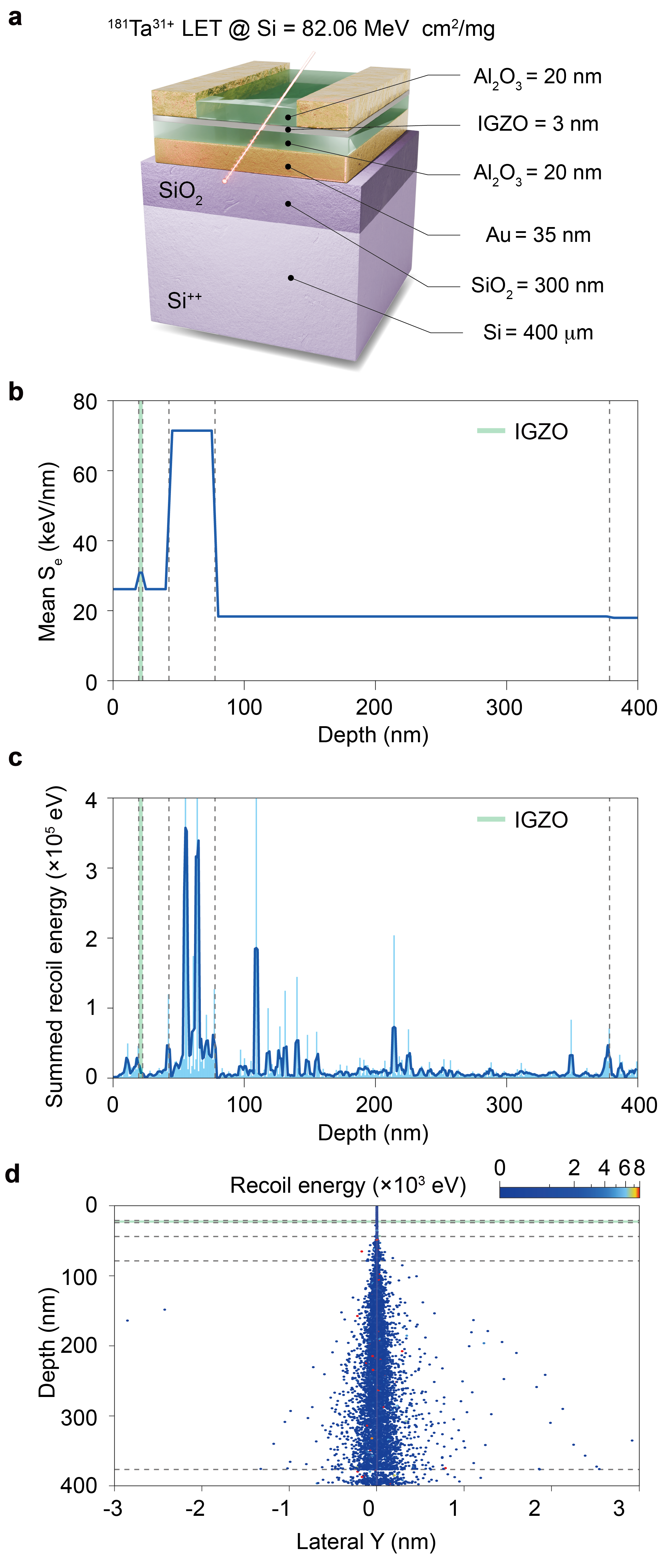}
 	\caption{\textbf{Simulated heavy-ion interaction with the ultrathin IGZO transistor structure.} \textbf{a}, Schematic of a $^{181}$Ta ions traversing the multilayer device stack. \textbf{b}, Electronic stopping power $S_{\mathrm{e}}$ as a function of depth, showing limited energy deposition in the $\sim$2 nm IGZO channel. \textbf{c}, Depth-dependent summed recoil energy, indicating reduced displacement damage in the IGZO layer compared to surrounding materials. \textbf{d}, Lateral distribution of collision events, showing highly localized damage along the ion track with minimal lateral spreading.
    }
 	\label{fig:fig4}
 \end{figure}

The transistor-level characteristics of the constituent devices are summarized in Fig. 1f-h. Figure 1f shows the transfer characteristics of the transistor used for the clock input path, exhibiting clear gate modulation and reliable switching behaviour. Figure 1g presents representative transfer curves of the MOS-FETs used to construct the inverter stages in the DFF, for different channel aspect ratios. These devices display well-defined field-effect behaviour over several orders of drain current modulation, confirming their suitability for logic integration.\cite{peng2024medium,fan2023two} The corresponding voltage transfer characteristics of a typical inverter are shown in Fig.~1h, where the output switches sharply over a narrow input-voltage window, indicating appreciable gain and robust logic-level restoration. The process and circuit-design leading to this operating window are detailed in Supplementary Note 1. Briefly, oxygen-plasma treatment and Al$_2$O$_3$ encapsulation improve the electrical response of the ultrathin IGZO channel, while systematic pull-up/pull-down transistor sizing tunes the inverter switching point and gain for DFF implementation (Supplementary Figs. 3-5). Figure 1i provides the statistical distribution of threshold voltage extracted from multiple IGZO transistors, which is important for evaluating device uniformity and the tolerance margin available for integrated circuit operation.

Finally, the circuit functionality of the DFF is verified in Fig. 1j. The frequency-dependent operation of the IGZO DFF was examined from 1 Hz to 1 kHz, confirming stable sequential switching over the tested range (Supplementary Fig. 6). Under a clock frequency of 1~kHz, the measured output signal follows the expected sequential switching behaviour, demonstrating successful data latching and release in the fabricated timing circuit. The stable operation of this DFF establishes that magnetron-sputtered amorphous IGZO transistors can support integrated sequential logic, providing the foundation for the subsequent investigation of radiation-resilient digital circuit operation under heavy-ion exposure.\cite{nomura2004room} Additional circuit demonstrations, including NAND, NOR, voltage-follower and four-stage shift-register operation, confirm that the same IGZO platform supports both combinational and sequential logic functions beyond the single-DFF unit (Supplementary Figs. 7-10 in Supplementary Note 2).

\vspace{5mm}
\noindent\textbf{DFF circuits made of IGZO FETs.} 
\\
To characterize the operation of the D-type flip-flops (DFFs) and the larger timing circuits derived from them, we developed a dedicated printed circuit board (PCB) platform and an accompanying electrical test configuration. A photograph of the experimental setup is shown in Fig.~2a, where a wave generator, power supply and computer-controlled readout system are connected to the sample under test. The custom PCB, shown in Fig.~2b, was designed to provide stable electrical interfacing to the thin-film chip while enabling programmable input, clock delivery and real-time signal acquisition. A detailed PCB-level readout architecture, including the sample interface, buffer stages,  microcontroller unit (MCU) and data-upload path, is provided in Supplementary Fig. 11. A photograph of the mounted chip is presented in Fig. 2c, and the corresponding measurement architecture is summarized in Fig. 2e. In this configuration, the hardware timing circuit delivers multiple output channels to the external electronics, while the input data and clock signals are synchronously supplied and monitored through the control and readout circuitry. This platform enables systematic evaluation of both single timing elements and cascaded sequential circuits.\cite{ao2025risc}

The basic circuit unit used in this work is the DFF introduced in Fig. 1. Each DFF adopts a master--slave configuration and contains two latching stages. In each stage, two inverters are combined with clock-controlled switching transistors to regulate the write and hold operations. As a result, one DFF consists of a total of 12 n-type IGZO transistors, including the inverter transistors and four clock-controlled field-effect transistors. By integrating these building blocks, we further constructed cascaded parallel-register circuits for sequential digital output. An optical micrograph of a representative cascaded DFF array is shown in Fig. 2d. In the chip presented here, 30 DFF units are integrated on a single circuit block, providing a sufficiently complex platform for evaluating multi-stage timing propagation and functional output generation.

Figure 2f shows a typical dynamic waveform measured from an individual DFF. Upon application of the clock signal, the output follows the expected sequential response to the input data, confirming correct latching and transfer behaviour. The stable phase relationship among the input, clock and output traces verifies that the amorphous IGZO transistors can support reliable edge-triggered timing operation over extended measurement intervals. After confirming the functionality of individual DFF units, we used eight such DFFs connected in series to construct a larger timing circuit for ASCII-coded digital output. This cascaded architecture forms the basis for the ``Hello World'' demonstration presented in the following figures\cite{hills2019modern}, and establishes the circuit complexity required for evaluating the radiation resilience of amorphous thin-film digital systems at the multi-device level.\cite{wachter2017microprocessor}

\vspace{5mm}
\noindent\textbf{Amorphous thin film timing circuits operating under \textit{in situ} heavy-ion irradiation.} 
\\
We next examined whether the amorphous thin-film timing circuits could maintain dynamic operation under \textit{in situ} heavy-ion irradiation. Before irradiation, cascaded DFF chains used for the “Hello World” hardware demonstration were monitored for continuous and reproducible operation under ambient conditions (Supplementary Fig. 12). Figure 3a shows a photograph of the powered irradiation experiment using $^{181}$Ta ions. The red laser beam was used only for sample alignment before the measurement and was switched off during irradiation and electrical testing. In this configuration, the circuit was continuously biased and clocked while the heavy-ion beam impinged on the chip, allowing direct evaluation of functional stability under simultaneous electrical operation and particle exposure.

We first tested the response of an individual DFF under irradiation. Figure~3b presents representative input, clock and output waveforms acquired while being exposed to $^{181}$Ta ion with the flux of $2.5\times10^{3}~$$\mathrm{ions\ cm^{-2}\ s^{-1}}$. In contrast to the non-irradiated operation shown in Fig. 2f, the DFF here was continuously operated during irradiation for 500 s. Despite the sustained heavy-ion exposure, the output retained the expected latching behaviour and followed the input sequence in synchrony with the clock, indicating that the basic timing element remained functional throughout the test duration. This result establishes that the sequential switching operation of the amorphous IGZO DFF is robust not only under ambient electrical testing but also during prolonged heavy-ion bombardment.

We then extended the evaluation from a single timing element to a larger cascaded circuit designed for ASCII-coded digital output. Figure 3c shows the input sequence of the 8-bit timing circuit, where the parallel digital states reproduce the programmed string ``HELLO WORLD'' in ASCII format. Figure 3d shows the corresponding output acquired under irradiation condition as input in Fig. 3b, namely continuous operation under $^{181}$Ta irradiation at $2.5\times10^{3}~$$\mathrm{ions\ cm^{-2}\ s^{-1}}$. The overall output sequence remains clearly recognizable as the intended ``HELLO WORLD'' string during the irradiation test, demonstrating that the multi-stage timing circuit preserves the programmed sequential logic under beam exposure. Within the 500 s measurement window, only a single erroneous code was observed, as marked by the red arrow in Fig. 3d. Consistent behaviour was observed in additional cascaded-register samples, which maintained the programmed “Hello World” output for a long operating time before rare DFF-level error events appeared (Supplementary Figs. 13 and 14 in Supplementary Note 3). The rarity of this error, compared with the otherwise correct string output over the full irradiation interval, highlights the high functional tolerance of the cascaded amorphous thin-film circuit to heavy-ion-induced disturbance.\cite{dodd2003basic} It is noted that the absolute circuit area of the present IGZO DFFs were substantially larger than that of advanced Si-CMOS flip-flops. A direct comparison of raw upset probability would therefore be dominated by layout footprint rather than intrinsic material and device response. To provide an auxiliary area-normalized assessment, we estimate the per-bit SEU cross section of the 8-bit IGZO DFF from the single observed upset event at a fluence of $1.0\times10^{6}$ ions/cm$^{2}$, yielding $\sigma_{\mathrm{SEU}}=1.25\times10^{-7}$ cm$^{2}$/bit. Using the extracted per-bit IGZO channel area, the corresponding area-normalized sensitivity factor is $\eta_A=\sigma_{\mathrm{SEU}}/A_{\mathrm{bit}}=3.075\times10^{-3}$, indicating that the effective SEU-sensitive cross section accounts for only $\sim$0.3075\% of the IGZO channel area. If this normalized sensitivity is projected onto a representative 22-nm silicon-on-insulator (SOI) technology, the equivalent SEU cross section is about $1.004\times10^{-11}$ cm$^{2}$/bit, which is smaller than the values typically reported for SOI-based circuits\cite{chi2022seu}. The detailed discussion is provided in Supplementary Note 4. Although this projection should not be interpreted as a strict device-scaling law, it shows that after correcting for the large experimental footprint, the ultrathin IGZO DFF exhibits low area-normalized SEU sensitivity under high-LET Ta-ion irradiation. This comparison further supports the advantage of ultrathin amorphous oxide channels for radiation-tolerant logic circuits. 

To further monitor the circuit state during beam exposure, we recorded the operating current as a function of time under the same irradiation condition. As shown in Fig. 3e, the current exhibits stable periodic modulation over 500 s, consistent with the repetitive switching of the timing circuit during sequential output generation. No catastrophic current collapse, runaway leakage or irreversible switching failure is observed during the measurement.\cite{soliman2007latchup,cecchetto2025energy} The circuit therefore sustains continuous operation up to a total fluence of $10^{6}$ ions/cm$^{2}$ under $^{181}$Ta irradiation with a linear energy transfer of 82 $\mathrm{MeV\,cm^{2}\,mg^{-1}}$. Together, these results demonstrate that amorphous thin-film timing circuits can preserve both transistor-level switching and system-level digital functionality during powered heavy-ion irradiation, supporting their potential for radiation-resilient electronics in harsh environments.

\vspace{5mm}
\noindent\textbf{Energy deposition and collision damage simulation of $^{181}$Ta ions in the IGZO FETs.} 
\\
To elucidate the physical origin of the experimentally observed radiation resilience, we performed particle-transport simulations (using SRIM and Geant4, see Methods) of heavy-ion interactions with the amorphous IGZO transistor structure.\cite{ziegler2010srim,agostinelli2003geant4} Figure 4a shows a schematic of a $^{181}$Ta ion traversing the multilayer device stack, including the Al$_2$O$_3$/IGZO/Al$_2$O$_3$/Au/SiO$_2$/Si structure. In the simulation, the full silicon substrate thickness (hundreds of micrometres) is included, while the plotted results focus on the near-surface region within the first $\sim$400 nm, where the active device layers are located. This allows direct evaluation of energy deposition and damage processes in the ultrathin IGZO channel. SRIM simulations further show that $^{181}$Ta ions exhibit a lower LET in IGZO than in Si under the same irradiation condition, indicating reduced ionization energy deposition in the active channel. Combined with the ultrathin IGZO geometry, this reduced LET further limits charge generation and helps mitigate single-event effects in the timing circuits (Supplementary Fig. 15 in Supplementary note 4).

The electronic stopping power ($S_{\mathrm{e}}$), which characterizes ion-induced electronic energy loss, is shown as a function of depth in Fig.~4b.\cite{sigmund2004stopping} As the $^{181}$Ta ion penetrates the multilayer structure, $S_{\mathrm{e}}$ exhibits layer-dependent variations reflecting differences in material composition and density. A local modulation of $S_{\mathrm{e}}$ is observed when the ion passes through the IGZO channel region, indicating that electronic energy deposition occurs within this active layer. However, because the IGZO channel thickness is only $\sim$2 nm, the total energy deposited within this region is intrinsically limited. As a result, the generation of transient charge and perturbation of the channel potential are strongly suppressed. This reduced energy deposition volume lowers the probability of inducing significant single-event transient effects, thereby minimizing the impact on the drain current and threshold voltage of the transistor.\cite{esposito2021investigating,martinella2020heavy}

In addition to electronic energy loss, we evaluated the nuclear collision processes associated with displacement damage. Figure 4c shows the depth-dependent distribution of accumulated recoil energy induced by $^{181}$Ta ions over the 0 - 400 nm region. Pronounced peaks in recoil energy are observed near the Au layer, where nuclear stopping is enhanced due to the higher atomic mass. In contrast, the IGZO channel region exhibits comparatively low recoil energy, indicating that direct displacement damage within this ultrathin layer is not the dominant process. Moreover, the extremely limited thickness of the IGZO channel constrains the volume available for atomic displacement, resulting in a negligible contribution of displacement damage within the active semiconductor layer, as further supported by depth-dependent vacancy-generation and collision-depth/recoil-energy analyses (Supplementary Figs. 16 and 17). These results suggest that heavy-ion-induced degradation in IGZO transistors is not governed by bulk damage accumulation in the channel.

The spatial distribution of collision events is further analysed in Fig. 4d, which maps the lateral spread of recoil events as a function of depth. Most collision events are confined within a narrow region along the ion trajectory, indicating minimal lateral scattering. Only a small fraction of high-energy recoil events extends laterally away from the primary track, reflecting localized collision cascades. This limited lateral dispersion implies that the damage induced by the incident ion remains highly localized and does not propagate across the channel region. For an ultrathin semiconductor layer such as IGZO, this localization further reduces the probability of large-area defect formation or collective electrical disruption.

Taken together, these simulations reveal that the ultrathin geometry of the amorphous IGZO channel plays a central role in suppressing both electronic and nuclear energy deposition effects. The combination of reduced interaction volume, limited displacement damage and confined spatial distribution of collision events provides a physical basis for the experimentally observed resilience of the timing circuits under heavy-ion irradiation. These results highlight ultrathin amorphous oxide semiconductors as an intrinsically robust platform for radiation-tolerant electronics.

\vspace{5mm}

\section*{Conclusion}

In summary, we have demonstrated amorphous ultrathin IGZO thin-film circuits capable of sequential digital operation under powered heavy-ion irradiation. Beyond individual device measurements, we realized functional sequential logic building blocks, including D-type flip-flops and cascaded parallel-register circuits, and further implemented an 8-bit timing system capable of outputting the ASCII-coded string ``HELLO WORLD''. This establishes that amorphous oxide semiconductors can support non-trivial digital operation at the circuit level, despite their structurally disordered nature. More importantly, the circuits remain operational under powered heavy-ion irradiation using $^{181}$Ta ions, retaining correct DFF switching and sustaining multi-bit ASCII output at a flux of $2.5\times10^{3}~$$\mathrm{ions\ cm^{-2}\ s^{-1}}$,  under a total fluence of $10^{6}$ ions/cm$^{2}$. This powered heavy-ion experiment provides, to our knowledge, the first circuit-level validation of IGZO sequential logic operation in a heavy-ion radiation environment. These results move radiation studies of amorphous semiconductors beyond static transistor characterization towards system-level demonstrations under realistic operating conditions. Together with the mechanistic understanding to be developed from simulation and comparative analysis, our work identifies ultrathin amorphous oxide electronics as a promising platform for lightweight, scalable and intrinsically radiation-resilient digital systems for harsh-environment applications.

\clearpage

 \clearpage

\section*{Methods}
\vspace{3mm}
\noindent\textbf{Fabrication of IGZO amorphous ultathin films.} Ultrathin IGZO films were deposited by magnetron sputtering using an AJA International ORION 8 system. A 99.99\%-pure IGZO target with an In$_2$O$_3$:Ga$_2$O$_3$:ZnO molar ratio of 1:1:1 was used for deposition. Prior to deposition, the chamber was evacuated to a base pressure below $1 \times 10^{-8}$ Torr. The sputtering was carried out in an Ar atmosphere at a flow rate of 33 sccm and a working pressure of 3 mTorr. The sputtering power was maintained at 50 W, and the substrate holder was rotated at 20 rpm to improve film uniformity.\cite{yabuta2006high} The deposition time was 2 min 40 s, yielding an ultrathin IGZO film with a thickness of approximately 2 nm.

\vspace{3mm}
\noindent\textbf{Fabrication of IGZO field-effect transistor and timing circuits.} 
First, bottom-gate patterns were defined on a SiO$_2$ substrate by photolithography, followed by metal deposition using electron-beam evaporation (EBE). The gate electrodes consisted of Ti/Au metal stacks with thicknesses of 5/30 nm. Subsequently, a 20 nm-thick Al$_2$O$_3$ dielectric layer was deposited by atomic layer deposition (ALD). To enable subsequent interlayer interconnection, via holes were formed in the dielectric layer by photolithographic patterning followed by CF$_4$-based reactive ion etching (RIE). The samples were then treated with oxygen plasma, followed by the deposition of an IGZO thin film by magnetron sputtering. After material deposition, the IGZO ultrathin layer was patterned by photolithography and etched using Ar-based RIE to remove the film outside the channel regions. Source/drain electrodes were fabricated on the patterned channel regions, together with additional interconnect electrodes. An Al$_2$O$_3$ layer serving as encapsulation and isolation layer was deposited by ALD. Contact holes were subsequently defined in the isolation layer for further interconnection. Finally, interconnect patterns were then formed by photolithography, followed by metal evaporation to deposit an additional Ti/Au interconnect electrode layer with thicknesses of 5/50 nm. 

Photolithography was carried out using an ABM/6/350/NUV/DCCD/SA mask aligner (ABM). Reactive ion etching was performed using a Samco RIE-10NR system. IGZO films were deposited by magnetron sputtering using an ORION 8 system (AJA International). Al$_2$O$_3$ layers were grown by ALD using a Savannah S200 system (Veeco). Metal electrodes were deposited using an Amod electron-beam evaporation system (Angstrom Engineering).

\vspace{3mm}
\noindent\textbf{Electrical measurements under ambient environment.} 
Initial electrical characterizations were performed on a probe station using a Cascade MPS150 system equipped with a Keithley 2400 source meter, a Keysight 33600A signal generator, and an Agilent B1500A semiconductor parameter analyzer. Once the samples were confirmed to exhibit the required current level, proper cascading behavior, and correct logic functionality, they were further tested on an MCU-based platform. In this setup, a 30-channel voltage-follower buffer circuit based on OPA4344 operational amplifier chips was designed to acquire output signals from the timing circuits samples, providing high input impedance, buffering, and electrical isolation. The buffered signals were then sampled by an STC32 microcontroller via analog-to-digital conversion (ADC). Real-time data acquisition, visualization, and dynamic plotting were implemented using host-computer software.

\vspace{3mm}
\noindent\textbf{Electrical measurements under radiation.} The electrical response of the timing circuits samples was characterized in situ during heavy-ion irradiation. The devices under test, denoted as S1, S2, S3 and S4, were designed based on FET devices using IGZO as the channel material. During the irradiation experiment, the samples were placed along the beam direction at a distance of approximately 2 cm from the Ta-ion beam exit window. Irradiation was performed using $^{181}$Ta ions. With Si as the reference material, the effective ion energy was 1334.92 MeV, corresponding to an effective LET$_{\mathrm{Si}}$ of 82.06 MeV$\cdot$cm$^{2}$/mg. During irradiation, the average ion fluence rate was maintained in the range of $2 \times 10^{3}$--$1 \times 10^{4}$ ions/cm$^{2}$/s.

Before irradiation, predefined input signals and clock signals were applied to each timing circuit to verify normal operation. Under these input and clock conditions, the timing circuits maintained a stable logic output of ``HELLO WORLD'', which was used as the reference output state for monitoring the radiation response. Throughout the irradiation process, the devices remained powered and functionally operational, enabling real-time observation of radiation-induced electrical disturbances under actual operating conditions.

The supply current and output response of the timing circuits were continuously monitored using an electronic measurement system. The measured electrical signals, including the output logic state and supply-current data, were recorded in real time by a computer. This measurement configuration enabled direct correlation between heavy-ion incidence events and changes in circuit behavior. In particular, the system was used to identify radiation-induced electrical anomalies, such as transient output errors, logic-state upsets, supply-current fluctuations, and possible single-event effects, including single-event upset (SEU) and single-event latch-up (SEL).

\vspace{3mm}
\noindent\textbf{Simulations.}
SRIM and Geant4 simulations were performed to investigate the interaction of energetic tantalum ions with the IGZO-based field-effect transistor stack and to evaluate the physical origin of its single-event-effect response. The simulated multilayer structure was constructed according to the experimental device geometry, consisting of 20 nm Al$_2$O$_3$, 2 nm IGZO, 20 nm Al$_2$O$_3$, 35 nm Au, 300 nm SiO$_2$, and 400 $\mu$m Si. A normally incident $^{181}$Ta ions beam with an energy of 10.30 MeV/u was considered. The corresponding effective LET in Si was 82.06 MeV$\cdot$cm$^{2}$/mg and was used as a reference high-LET irradiation condition.

SRIM calculations were first carried out to estimate the depth-dependent electronic stopping, nuclear stopping, recoil generation, and vacancy production in the multilayer device stack. The simulated ion tracks and recoil distributions were used to identify the regions where ion-induced displacement damage and secondary collision events are most likely to occur. Although the IGZO channel is located in the shallow region of the device stack, its ultrathin thickness of only $\sim$2 nm substantially reduces the effective sensitive volume exposed to the incident heavy ions. Consequently, the total energy deposited within the IGZO channel and the corresponding ion-induced charge generation are geometrically constrained, even under high-LET Ta-ion irradiation.

\section*{\label{sec:level1}Data Availability}

The data that support the findings of this study are available via Zenodo at ...
%\textcolor{blue}{https://doi.}

\section*{\label{sec:level2}Code Availability}

The code that support the findings of this study are available upon reasonable request to the corresponding authors.

\section*{\label{sec:level3}Acknowledgements}
This work is supported by the National Key R$\&$D Program of China (No. 2022YFA1203903 and No.2021YFA1601300) and the National Natural Science Foundation of China (NSFC) (Grant Nos. 92265203, 11974357, U1932151, 12204287, and 11974027). Z.V.H. acknowledges the support of the Fund for Shanxi “1331 Project” Key Subjects Construction, and the Innovation Program for Quantum Science and Technology (grant no. 2021ZD0302003). H. W. acknowledges the support of National Natural Science Foundation of Liaoning Province, China (No. 2024JH3/5010 0 022 ).

\section*{Author Contributions}
J.L., Z.H., H.W. and Y.Y. conceived the experiment and supervised together with L.L. and C.Z. the overall project. X.S., Y.P. and W.W. performed the IGZO synthesis and device fabrications. X.S., S.L., C.Z., J.L. and L.L. performed electrical measurements under radiations; J.L., Z.H., X.L., S.L., H.W. and X.S. analyzed the experimental data. The manuscript was written by J.L., Z.H., X.L., H.W., X.S. and S.L. with discussion and inputs from all authors.

\section*{Competing Interests}
The authors declare no competing interests.

\end{document}